\documentclass[epj]{svjour}
\usepackage{amssymb}
\usepackage{amsmath}
\usepackage{graphicx}

\newcommand{\ie}[0] {{\it i.e.},\ }
\newcommand{\eg}[0] {{\it e.g.},\ }
\newcommand{\KT}[0] {k_{\rm B}T}
\newcommand{\Half}[0]{\frac{1}{2}}
\newcommand{\D}[2]{\frac{\partial #1}{\partial #2}}

\newcommand{\Fig}[1] {Fig.\ \ref{#1}}
\newcommand{\Eq}[1] {Eq.\ (\ref{#1})}
\newcommand{\Eqs}[2] {Eqs.\ (\ref{#1}) and (\ref{#2})}

\newcommand{\A}[0]{\langle A \rangle}
\newcommand{\Amax}[0]{A_{\rm max}}
\newcommand{\Pc}[0]{p_{\rm c}}
\newcommand{\RgSq}[0]{\langle R_{\rm g}^2 \rangle}
\newcommand{\R}[0]{{\bf r}}

\begin{document}

\title{Smoothening Transition of a Two-Dimensional Pressurized Polymer Ring}

\author{Emir Haleva \and Haim Diamant}

\institute{School of Chemistry, Raymond \& Beverly Sackler Faculty of
  Exact Sciences, Tel Aviv University, Tel Aviv 69978, Israel}

\date{Received: date / Revised version: date}

\abstract { We revisit the problem of a two-dimensional polymer ring
    subject to an inflating pressure differential.  The ring is
    modeled as a freely jointed closed chain of $N$ monomers. Using a
    Flory argument, mean-field calculation and Monte Carlo
    simulations, we show that at a critical pressure, $\Pc\sim
    N^{-1}$, the ring undergoes a second-order phase transition from a
    crumpled, random-walk state, where its mean area scales as $\A\sim
    N$, to a smooth state with $\A\sim N^2$. The transition belongs to
    the mean-field universality class. At the critical point a new
    state of polymer statistics is found, in which $\A\sim
    N^{3/2}$. For $p\gg \Pc$ we use a transfer-matrix calculation to
    derive exact expressions for the properties of the smooth state.
 }

\maketitle
\section{Introduction}  \label{secIntroduction}

Considerable theoretical efforts were directed during the 1980s and
1990s at random polymer rings constrained to a plane, both as a
fundamental problem of statistical mechanics
\cite{Butler1987,Wiegel1988,Duplantier1989,Duplantier1990,Cardy1994}
and as a highly idealized model for membrane vesicles
\cite{Fisher1987,Fisher1990,Fisher1990_2,Rudnick1991,Rudnick1993,Levinson1992,Maritan1993,Gaspari1993,Vilgis2005}.
In addition, there was much interest in transitions between crumpled
and smooth states of membranes \cite{Kantor1987,Peliti1988}.

Two bodies of work, in particular, concerned pressurized
two-dimensional (2D) rings. Fisher {\it et al.} studied both inflated
and deflated, closed, 2D self-avoiding walks using Monte Carlo
(MC) simulations and scaling analysis
\cite{Fisher1987,Fisher1990,Fisher1990_2}.  For a finite ring of $N$
monomers, subject to an inflating pressure differential $p>0$, three
regimes were found:
(i) a weak-inflation crumpled regime for $0<pN^{2\nu}\lesssim 1$ ($\nu=3/4$
being the swelling exponent of 2D self-avoiding walks), where the mean area scales as
$\A\sim N^{2\nu}f_A(pN^{2\nu})$;
(ii) a strong-inflation crumpled regime for $1\lesssim pN^{2\nu} \ll N^{2\nu-1}$,
where the same scaling holds but the scaling function $f_A$ becomes a power law; and
(iii) a smooth regime for $pN^{2\nu}\gtrsim N^{2\nu-1}$, where $\A\sim
N^2$.
The crossovers between these regimes are gradual with no phase
transitions. In the thermodynamic limit, defined as
$N\rightarrow\infty$ and $p\rightarrow 0$ such that $pN^{2\nu}$ is
finite, the range of the scaling regime (ii) in the case of
self-avoiding walks ($\nu=3/4$) becomes infinite. Note, however, that
in the random-walk case ($\nu=1/2$) this regime disappears; we will
show below that for random walks regime (ii) turns, in fact, into a
second-order phase transition between regimes (i) and (iii).

The second body of works consists of analytical studies of closed,
pressurized, 2D random walks, first presented by Rudnick and Gaspari
\cite{Rudnick1991,Rudnick1993,Levinson1992}. In this model the
polymer comprises a set of Gaussian springs of fixed elastic constant
and no bending rigidity.  At a critical pressure $\Pc\sim N^{-1}$ the
mean area was found to diverge (\ie the ring inflates to
infinity). This divergence, obviously, is made possible by the
extensibility of the chain, \ie the ability of the springs in this
model to be infinitely stretched.

In the current work we revisit the Rudnick-Gaspari model while
imposing inextensibility of the chain, as is actually appropriate for
real polymers or vesicles. This constraint changes the infinite
inflation at $p=\Pc$ into a second-order phase transition between a
crumpled, random-walk state and a smooth one.
The model is defined in Section \ref{secModel}. We begin the
investigation in Section \ref{secFlory} with a simple Flory argument
which, nevertheless, yields the correct physics. In Section
\ref{secMF} we present a mean-field calculation which accurately
captures the behavior below and at the transition. In Section
\ref{secTM} we derive exact expressions for the behavior at $p\gg\Pc$
using a transfer-matrix technique. The behavior found in Sections
\ref{secFlory}-\ref{secTM} is verified in Section \ref{secMC} by MC
simulations. Finally, the results are summarized and discussed in
Section \ref{secSummary}.

\section{Model}\label{secModel}

\begin{figure}[tbh]
  \centering
  \vspace{1cm}
  \includegraphics[height=2.2in]{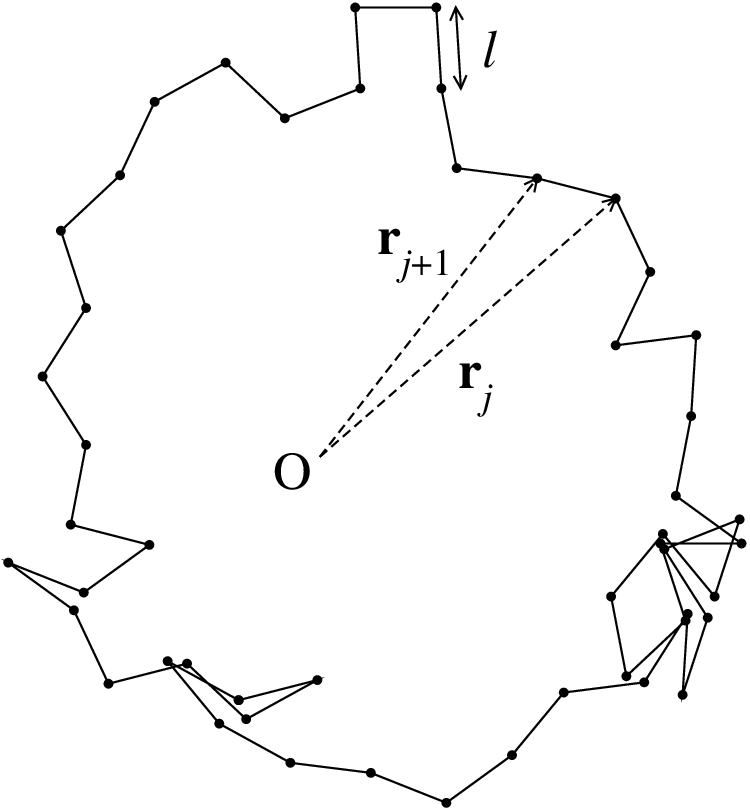}
  \caption{Schematic illustration of the ring and its parameters.}
  \label{figModel}
\end{figure}

The system under consideration is illustrated in \Fig{figModel}.  A
closed, two-dimensional, freely jointed chain of $N$ monomers is
subject to an inflating 2D pressure differential $p>0$ between its
interior and exterior. The monomers are connected by rigid links of
length $l$. We define $l\equiv 1$ as the unit length and the thermal
energy $\KT$ as the unit energy. (Thus, $p$ is measured in units of
$\KT/l^2$.)  The chain is ideal, \ie there are no interactions between
monomers. (Effects related to self-avoidance will be briefly discussed
in Sections \ref{secFlory} and \ref{secSummary}.) In addition, no
bending rigidity is taken into account, \ie the chain is freely
jointed.  A configuration of the ring is defined by a set of 2D
vectors $\{\R_j\}_{j=0\ldots N}$ specifying the positions of the
monomers. The condition that the chain be closed is expressed by
${\R_0} = {\R_N}$.

The probability of a specific configuration is
\begin{equation}
  \label{eqProbability}
  P\left(\{\R_j\},p\right) \propto e^{p A[\{\R_j\}]} \prod_{j=1}^{N} \delta(|\R_{j}-\R_{j-1}|-1),
\end{equation}
where $A$ is the area enclosed by the ring.  As in previous works
\cite{Wiegel1988,Duplantier1989,Rudnick1991,Rudnick1993,Levinson1992},
we take $A$ as the {\it algebraic} area rather than the
geometrical one,
\begin{equation}
\label{eqAlgebraicArea}
 A[\{\R_j\}]=\Half\sum_{j=1}^{N} (\R_{j-1}\times\R_j)\cdot\hat{\bf z},
\end{equation}
where $\hat{\bf z}$ is a unit vector perpendicular to the plane of
the ring. This area may take both positive and negative values. At
zero pressure both signs are equally favorable and the mean area must
vanish. At high pressures the probability of configurations with
negative $A$ is exponentially small in $p|A|$, and whether one takes
the algebraic or geometrical area will become statistically
insignificant. (We shall further discuss this assumption in Section
\ref{secSummary}.)

Using Eqs.\ (\ref{eqProbability}) and (\ref{eqAlgebraicArea}), we
write the partition function of the ring as
\begin{equation}
  \label{eqPartitionFunction}
  Z(p,N) = \int \prod_{j=1}^{N} d\R_j e^{\Half p(\R_{j-1}\times\R_j)\cdot\hat{\bf z}} 
  \delta(|\R_{j}-\R_{j-1}|-1),
\end{equation}
with $\R_0=\R_N$.

\section{Flory Argument}\label{secFlory}

We begin the analysis with a simple Flory argument
\cite{PolymerPhysics2003} which captures most of the physics to be
more rigorously treated in the following sections.  The free energy of
the ring (in units of $\KT$) is expressed as a function of $R$, the
radius of the statistical cloud of monomers (\ie the root-mean-square
radius of gyration). We divide it into three terms,
\begin{eqnarray}
  \label{eqFloryEnergy}
  F(R) &=& F_{\rm el} + F_{\rm inext} + F_p \nonumber\\
  F_{\rm el} &\sim& R^2/N \nonumber\\
  F_{\rm inext} &\sim& R^4/N^3 \nonumber\\
  F_p &\sim& -pR^2.
\end{eqnarray}
The elastic term, $F_{\rm el}$, is the usual entropic-spring free
energy of a Gaussian chain \cite{PolymerPhysics2003}. The second term
is the leading non-Gaussian correction due to the inextensibility of
the chain [see Appendix, \Eq{eqApF}], needed here to stabilize the
ring against infinite expansion. The last term is the pressure
contribution, where the mean area of the ring is taken as proportional
to $R^2$ \cite{Duplantier1990,Cardy1994}.

Equation (\ref{eqFloryEnergy}) has the form of a Landau free energy,
describing a second-order transition at $p=\Pc\sim N^{-1}$.  Since the
critical pressure depends so strongly on system size, we use hereafter
the rescaled pressure $\hat p\equiv p/\Pc\sim p N$, and define the
thermodynamic limit as $N\rightarrow\infty$ and $p\rightarrow0$ such
that $\hat p$ is finite. For $\hat p<1$ $F_{\rm inext}$ is negligible,
and $R$ has a Gaussian distribution with $\langle R^2\rangle \sim
N(1-\hat p)^{-1}$. For $\hat p>1$ we have $R^2\sim N^2(\hat p-
1)$. Thus, defining an order parameter $M=R/N$, we find in the
thermodynamic limit
\begin{equation}
\label{eqFloryM}
M \sim \begin{cases} 0 \ \ \ & \hat p<1\\ 
    (\hat p-1)^{\beta},\ \ \beta=1/2 \ \ & \hat p>1.
  \end{cases} 
\end{equation}
At the critical point itself $F=F_{\rm inext}\sim R^4/N^3$, and $R$
has a non-Gaussian distribution with 
\begin{equation}
  \label{eqFloryRgPc}
  \langle R^2(\hat p=1)\rangle \sim N^{2\nu_{\rm c}}, \ \ \nu_{\rm c} =3/4.
\end{equation}

Note that the competition between $F_{\rm el}$ and $F_p$, leading to
the second-order transition, is unique to 2D. In addition, when an
excluded-volume term, $F_{\rm ev} \sim N^2/R^2$, is added to the free
energy, \Eq{eqFloryEnergy}, the transition is removed. This agrees
with previous studies of self-avoiding rings
\cite{Fisher1987,Fisher1990,Fisher1990_2}, which did not report any
phase transition upon increasing pressure.

\section{Mean-Field Theory}\label{secMF}

\begin{figure}[tbh]
  \centering
  \vspace{0.4cm}
  \includegraphics[width=3.2in]{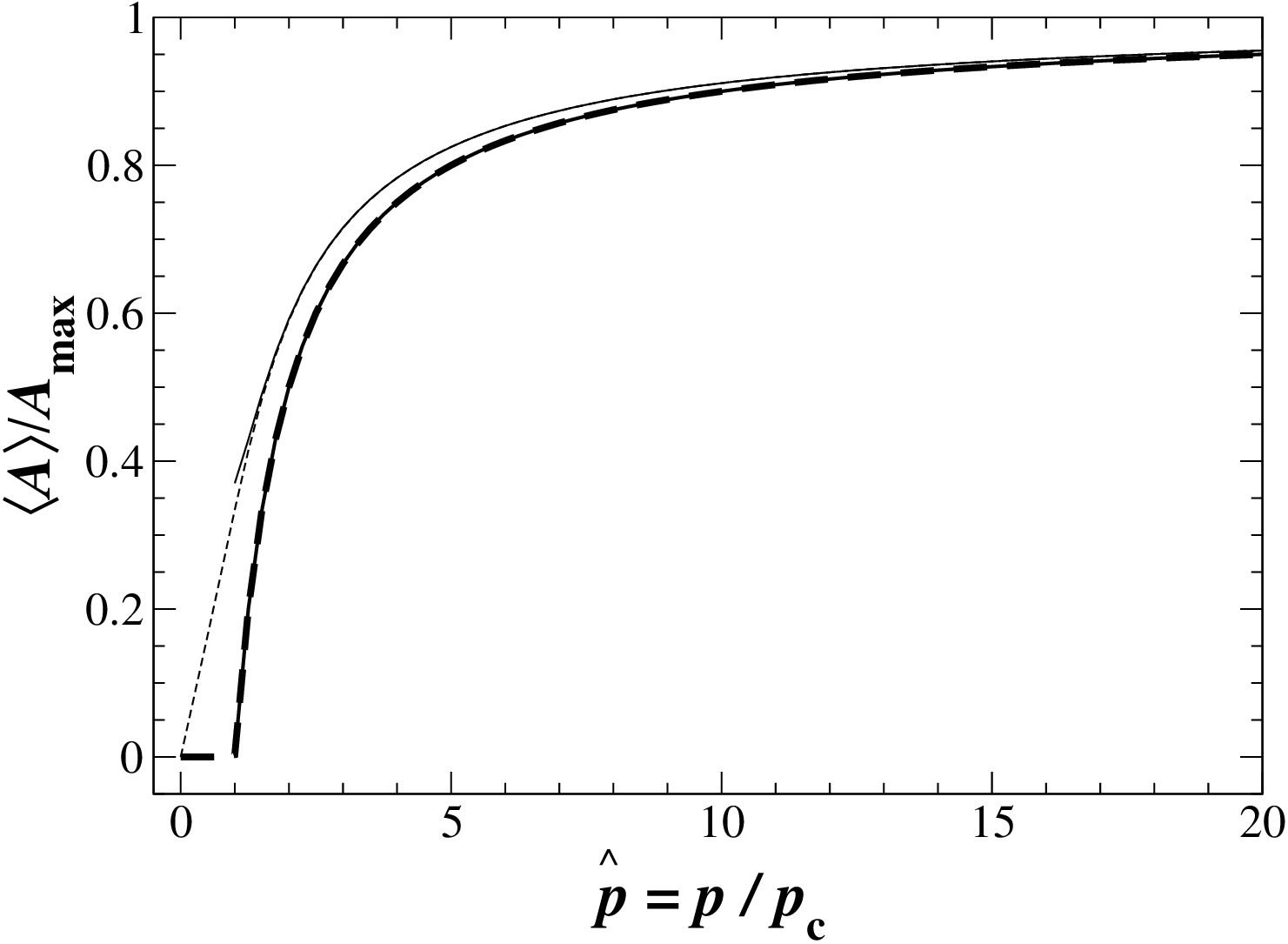}
  \caption[Mean area dependence on pressure]{The mean area in units of
    $A_{\rm max}\sim N^2$ as a function of the rescaled pressure (in
    units of $\Pc \sim N^{-1}$) as obtained from the mean-field
    approximation. The dashed curves are calculated numerically using
    \Eqs{eqMeanUSquare}{eqMeanArea2}, whereas the solid curves
    (calculated only for $\hat p \geq 1$) present the approximation
    given by \Eq{eqMeanArea3}.  Calculations were performed for $N=10$
    (thin curves) and $N=10^5$ (thick curves). }
  \label{figAreaMF}
\end{figure}

In this section we calculate the partition function of the freely
jointed ring, \Eq{eqPartitionFunction}, by relaxing the rigid
delta-function constraints on link lengths into harmonic
potentials. The spring constant $\lambda$ of the links is chosen so as
to make the root-mean-square length of a link equal to $l=1$. This
type of approximation, first suggested by Harris and Hearst
\cite{Harris1966}, was successfully employed in studies of the
Karatky-Porod worm-like-chain model \cite{Harris1966,Thirumalai1995},
where it was shown to be equivalent to a mean-field assumption (for
the field conjugate to the rigid link-length constraints).  The
partition function contains now only Gaussian terms and, therefore,
can be calculated exactly,
\begin{equation}
  \label{eqMFPartitionFunction}
  \begin{split}
  Z(p,N,\lambda) &= \int \prod_{j=1}^{N} d\R_j e^{\Half
  p(\R_{j-1}\times\R_j)\cdot\hat{\bf z} - \lambda (\R_j-\R_{j-1})^2} \\
  &=\frac{1}{\lambda^N}\frac{N p}{4 \lambda \sin{(\frac{N p}{4
  \lambda})}}.
  \end{split}
\end{equation}
(The spring constant $\lambda$ is in units of $\KT/l^2$.)  Details of
the calculation can be found in Ref. \cite{Rudnick1993}. This result
can be obtained also by analogy to the quantum propagator of a charged
particle in a magnetic field \cite{FeynmanHibbs}.  
The mean area is obtained by differentiation with respect to $p$,
\begin{equation}
  \label{eqMeanArea}
  \langle A(p,N,\lambda) \rangle = \D{\ln Z}{p} = 
  \frac{1}{p} - \frac{N \cot{(\frac{N p}{4\lambda})}}{4 \lambda}.
\end{equation}
For $\lambda=1$ \Eq{eqMeanArea} is the same as the result obtained by
Rudnick and Gaspari \cite{Rudnick1991,Rudnick1993,Levinson1992},
exhibiting a divergence at
\begin{equation}
  \label{eqPc}
  \Pc=4\pi/N.  
\end{equation}
Yet, in our case $\lambda$ is not fixed but is to be determined
self-consistently to ensure the softened inextensibility constraint.
It is clear that, as the pressure increases, the springs must become
stiffer to satisfy this constraint.
To impose the constraint, we calculate the mean-square link length and
set it to 1,
\begin{equation}
  \label{eqMeanUSquare}
  \begin{split}
    \langle (\R_j-\R_{j-1})^2 \rangle &= -\frac{1}{N} \D{\ln Z}{\lambda}\\
    &= \frac{1}{\lambda} + \frac{p}{N\lambda}\left(\frac{1}{p} - \frac{N \cot{(\frac{N
    p}{4\lambda})}}{4 \lambda}\right) = 1,
  \end{split}
\end{equation}
thus obtaining a transcendental equation for $\lambda(p,N)$.
Equations (\ref{eqMeanArea}) and (\ref{eqMeanUSquare}) are combined to
yield a simpler expression for $\A$ as a function of $\lambda$,
\begin{equation}
  \label{eqMeanArea2}
  \langle A(p,N,\lambda)\rangle = \frac{N(\lambda - 1)}{p}.
\end{equation}
Numerical solution of \Eq{eqMeanUSquare} for $\lambda$ [in the range
$\lambda > Np/(4\pi)$] and substitution of the result in
\Eq{eqMeanArea2} yield the mean area as a function of $p$ and $N$.
(See dashed curves in \Fig{figAreaMF}.)

For very low pressures, $p\ll \Pc$, we expand \Eq{eqMeanArea} to first
order in $p$ to get
\begin{equation}
  \label{eqLowMeanArea}
  \langle A(p\ll\Pc,N)\rangle =\frac{1}{48} N^2 p,
\end{equation}
\ie a linear dependence on $p$ as expected from linear
response. (Recall that at $p=0$ the mean algebraic area vanishes.)

For higher pressure we obtain a good approximation for $\lambda(p,N)$
in the limit $N\gg 1$ by substituting in \Eq{eqMeanUSquare}
$\cot[N p/(4\lambda)] \simeq [N p/(4 \lambda) - \pi]^{-1}$.
This gives
\begin{equation}
  \label{eqLambda}
  \lambda(\hat{p},N\gg 1) \simeq \frac{\hat{p}+1+\frac{1}{N} + \sqrt{(\hat{p}-1)^2 +\frac{2}{N}
  (\hat{p}+1)+\frac{1}{N^2}}}{2},
\end{equation}
where $\hat{p}\equiv p/\Pc = pN/(4\pi)$ is the rescaled pressure.  In
 the thermodynamic limit \Eq{eqLambda} reduces to the continuous but
 nonanalytic function
\begin{equation}
\label{eqLambdaLimit}
\lambda(\hat{p},N\rightarrow\infty) = \begin{cases}
1 & \hat{p}<1\\
\hat{p} & \hat{p}>1.
\end{cases}
\end{equation}
Equation (\ref{eqLambda}) should be regarded as an asymptotic
expression in the limit $N p/(4 \lambda)\rightarrow\pi$, which turns
out to be valid for any $p\not\ll\Pc$. (A Taylor expansion around this
point fails because of the nonanalyticity inferred above.)
Substituting \Eq{eqLambda} in \Eq{eqMeanArea2} yields an approximate
expression for $\A$ as a function of $\hat p$ and $N,$
\begin{equation}
  \label{eqMeanArea3}
  \begin{split}
    &\langle A(\hat{p}\not\ll 1,N\gg 1)\rangle \simeq \\ 
    &\frac{N^2}{4\pi}\frac{\hat{p}-1+\frac{1}{N} + \sqrt{(\hat{p}-1)^2
    +\frac{2}{N}(\hat{p}+1)+\frac{1}{N^2}}}{2\hat{p}}.
  \end{split}
\end{equation}

In the thermodynamic limit, the behavior of $\A$ around and above the critical pressure
is obtained from \Eq{eqMeanArea3} as 
\begin{equation}
  \label{eqMeanAreaAtPc}
  \A =\begin {cases} 
  \frac{N}{4\pi}\frac{1}{\hat{p}(1-\hat{p})} \xrightarrow{\hat{p}\rightarrow 1^-}
  \frac{N}{4\pi}\frac{1}{1-\hat{p}}&1-\hat{p} \gg N^{-1/2}\\
  \frac{N^{3/2}}{4\pi} & |1-\hat{p}|\ll N^{-1/2}\\
  \frac{N^2}{4\pi}\frac{\hat{p} - 1}{\hat{p}} \xrightarrow{\hat{p}\rightarrow 1^+}
  \frac{N^2}{4\pi}(\hat{p} - 1)  &\hat{p}-1 \gg N^{-1/2},
  \end{cases}
\end{equation}
revealing a continuous (second-order) transition. Below the transition
we get the same behavior as in the Rudnick-Gaspari model
\cite{Rudnick1991,Rudnick1993}, $\A\sim N(1-\hat{p})^{-1}$. Yet, due
to the inextensibility in our model, the increase of $\A$ as
$\hat{p}\rightarrow 1^-$ breaks at $|1-\hat{p}|\sim N^{-1/2}$.  In the
transition region, $|1-\hat p|\ll N^{-1/2}$, which shrinks to a
point in the thermodynamic limit, we find $\A\sim N^{2\nu_{\rm
c}},\nu_{\rm c} = 3/4$. Above the transition the ring reaches a smooth
state with $\A\sim N^2(\hat p -1)/\hat p$. All of these results agree with
the findings of the Flory argument presented in Section \ref{secFlory}
once we identify $\A\sim\RgSq$. As $p$ is increased to infinity, $\A$
tends, as it should, to
\begin{equation}
  \label{eqMaxMeanArea}
  \langle A(\hat{p}\rightarrow\infty,N)\rangle=\Amax=\frac{N^2}{4\pi},
\end{equation}
which is the area of a circle of perimeter $N$.

Figure \ref{figAreaMF} shows the dependence of $\A$ on $\hat{p}$ for
$N=10$ and $10^5$ calculated both from the numerical solution of
\Eqs{eqMeanUSquare}{eqMeanArea2}, and using the approximate expression
(\ref{eqMeanArea3}). For large $N$ the critical behavior becomes
apparent, with a transition between two distinct states---one in which
$\A\sim N$ and, hence, in units of $A_{\rm max} \sim
N^2$, the mean area vanishes for $N\rightarrow\infty$, and another
with a mean area proportional to $\Amax\sim N^2$. As can be seen in
\Fig{figAreaMF}, the approximate expression (\ref{eqMeanArea3}) is
practically indistinguishable from the numerical solution for $\hat p
\gtrsim 1$, even for small $N$.

The compressibility (defined with respect to the reduced pressure
$\hat p$) is obtained from \Eq{eqMeanAreaAtPc} as

\begin{equation}  
  \label{eqCompressibility}  
  \kappa = \frac{1}{A} \D{A}{\hat{p}} \xrightarrow{\hat{p}\rightarrow 1^\pm} |\hat{p} -1|^{-\gamma}, \ \  \gamma=1.
\end{equation}
At the critical point itself the compressibility diverges as 
\begin{equation}
  \label{eqMeanCompAtPc}
  \kappa(\hat{p}=1) = \Half N^{1/2}.
\end{equation}

To calculate the mean-square radius of gyration, $\RgSq=N^{-1}\sum
\R_j^2$ (where $\R_j$ are measured with respect to the center of
mass), we add a $h \sum \R_j^2$ term to the Hamiltonian of
\Eq{eqMFPartitionFunction} and differentiate the resulting partition
function with respect to $h$. This yields
\begin{equation}
  \label{eqMeanRg}
  \begin{split}
    \langle R_{\rm g}^2(p,N,\lambda)\rangle &= \left. \frac{1}{N}\D{\ln{Z(p,N,\lambda,h)}}{h} \right|_{h=0} =\\
    &= \frac{4\lambda - N p \cot{(\frac{N p}{4\lambda})}}{N p^2} = \frac{4\lambda}{N p} \A{}.
  \end{split}
\end{equation}
For $\hat{p} \ll 1$, combining this result with
\Eqs{eqLowMeanArea}{eqLambdaLimit}, we get $\RgSq = N/12$, which is
the well known result for the mean-square radius of gyration of a
Gaussian ring (\eg \cite{PolymerPhysics2003}). For large pressures,
$\hat p\geq 1$, we have from \Eq{eqLambdaLimit}
$\lambda\xrightarrow{N\rightarrow\infty}\hat{p}= p N/(4\pi)$, thereby
recovering the relation for an average circle, $\pi\RgSq =\A$.

\begin{figure}[tbh]
  \centering
  \vspace{0.5cm}
  \includegraphics[width=3.2in]{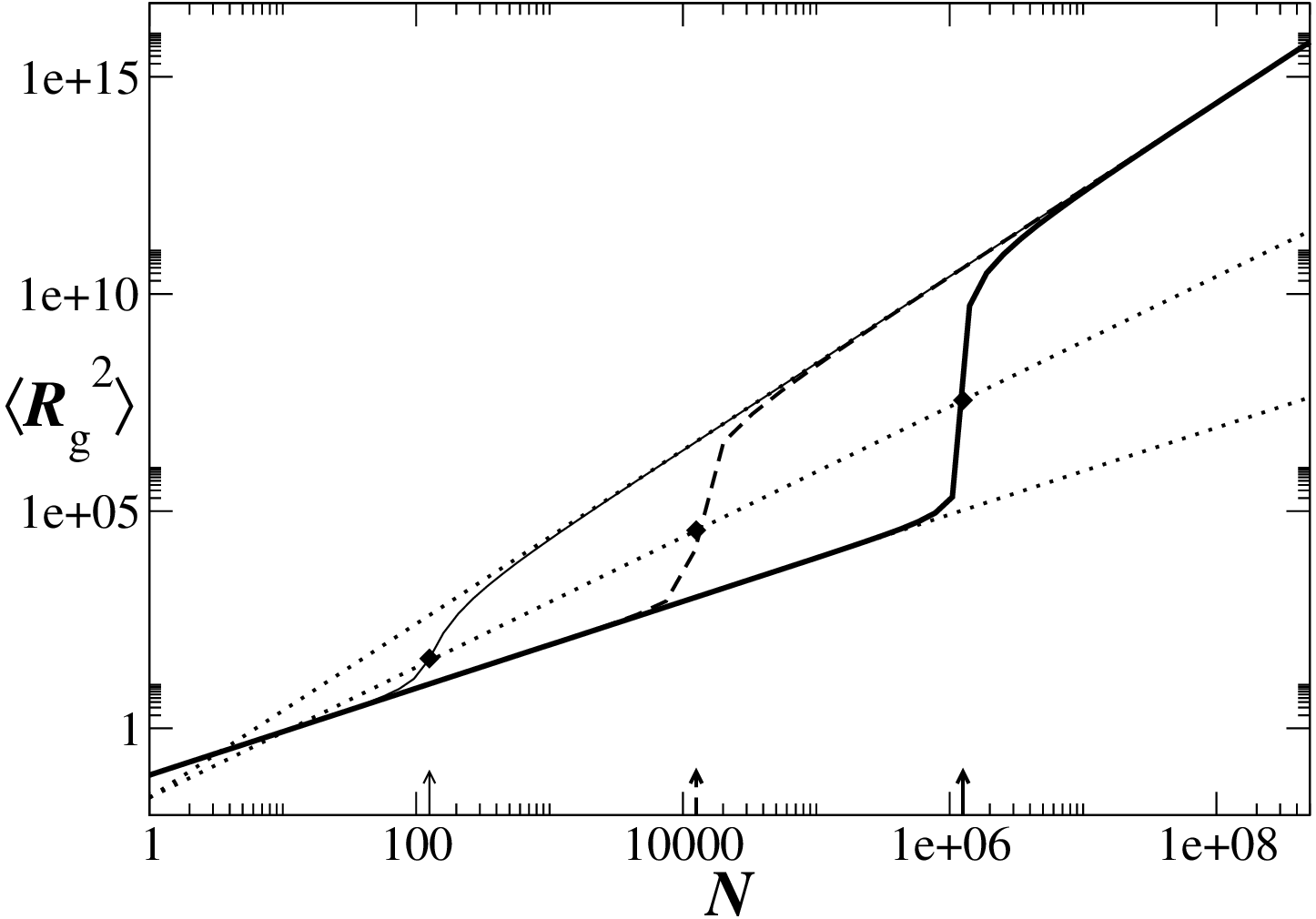}
  \caption[Mean-square radii of gyration dependence on n]{Mean-square
     radius of gyration as a function of $N$ at fixed values of
     $p=10^{-5}, 10^{-3}$ and 10$^{-1}$ (thick, dashed and thin
     curves, respectively) as obtained from the mean-field theory.
     For $N<N_c=4\pi/p$ (marked with arrows) $\RgSq \sim N$, while for
     $N > N_c$ $\RgSq \sim N^2$.  At the critical point (diamonds)
     $\RgSq$ is proportional to $N^{3/2}$.  The dotted lines show,
     from bottom to top, the dependencies for an unpressurized ring
     ($\RgSq =N/12$), for a ring at the critical pressure [$\RgSq
     =N^{3/2}/(4\pi^2)$] and for a stretched circle [$\RgSq
     =N^2/(4\pi^2$)]. }
  \label{figRgMF}
\end{figure}

Figure \ref{figRgMF} shows the dependence of $\RgSq$ on $N$ at fixed
pressure. The data were obtained by substituting the numerical
solution for $\lambda$ [\Eq{eqLambda}] in \Eq{eqMeanRg}. The scaling
of $\RgSq$ changes at the critical point $N_{\rm c}=4\pi/p$. Below the
critical point, $N<N_{\rm c}$, $\RgSq \sim N$, as in a Gaussian
chain. Above it $\RgSq \sim N^2$, as in a stretched chain.  As $p$
decreases, the transition becomes sharper. At exactly $N=N_{\rm c}$
$\RgSq$ scales as $N^{3/2}$. Thus, the analysis of $\RgSq$ yields the
same scaling with $N$ as obtained for $\A$.

\section{Transfer-Matrix Formulation}\label{secTM}

Since the interactions between all adjacent monomers are identical,
including the one between the first and the $N$th, the partition
function may be rewritten in the following transfer-matrix form:
\begin{equation}
  \label{eqExactPartitionFunction}
  \begin{split}
    &Z(p,N) = \underset{\{\R_j\}}{\rm Tr} \prod_{j=1}^N T(\R_{j-1},\R_j),\ \ 
    \R_0=\R_N \\
    &T(\R,\R')= e^{\frac{p}{2}(\R\times\R')\cdot\hat{\bf z}}\delta(|\R-\R'|-1).
  \end{split}
\end{equation}
Solution of the associated eigenvalue problem,
\begin{equation}
  \label{eqTOperator}
  \mu \Psi(\R)=\int d\R' T(\R,\R') \Psi(\R'),
\end{equation}
(in particular, finding the two eigenvalues $\mu$ of largest absolute
value) will yield the exact solution of the model.

Two properties of the operator $T$ are readily noticed: it is
non-Hermitian, $T(\R,\R')\neq T(\R',\R)$, and it is
rotational-invariant.  To exploit the invariance to rotations, we
change to polar coordinates, $\R= (\rho, \varphi$), and separate
variables as $\Psi(\R) = \chi(\rho)e^{i m \varphi}$, where $m=0,\pm
1,\pm 2,\ldots$ to maintain periodicity in $\varphi$. Equation
(\ref{eqTOperator}) can then be integrated over angles to give
\begin{equation}
  \label{eqROperator}
  \begin{split}
    &\mu \chi(\rho) = \int_{\rho-1}^{\rho+1} d\rho' \tilde{T}(\rho,\rho')\chi(\rho')\\ 
    &\tilde T(\rho,\rho')=\frac{2\cosh\left(p\rho\rho'\sqrt{1-\Delta^2}/2+ im\cos^{-1}\Delta\right)}{\rho\sqrt{1-\Delta^2}}\\
    &\Delta(\rho,\rho')=\frac{\rho^2 +\rho'^2 -1}{2\rho \rho'},
  \end{split}
\end{equation}
thus reducing the original operator to the one-dimensional operator
$\tilde{T}$. 

Unfortunately, we have not been able to diagonalize $\tilde T$ for any
$p$. For $p\gg\Pc$, nevertheless, the ring has stretched
configurations, and we can assume that the distances of all monomers
from the center of mass are much larger than the link length, $\rho\gg
1$. We expand $\chi(\rho')$ around $\rho$ to zeroth order, change
variables according to $\rho'=\rho+\sin\theta$, and integrate over
$\theta$ to get
\begin{equation}
  \label{eqDTOperator}
  \mu^{(0)} \chi(\rho) = 2 \pi I_0(p \rho /2)\chi(\rho),
\end{equation}
where $I_0$ is the zeroth-order modified Bessel function of the first
kind. Thus, at this order of approximation, the eigenfunctions have
the form $\chi_k^{(0)}(\rho)= [N/(2\pi
\rho_k)]^{1/2}\delta(\rho-\rho_k)$ with a continuous spectrum of
eigenvalues. The spectrum is bounded from above by $\mu^{(0)}_{\rm
max} = 2 \pi I_0(p \rho_{\rm max}/2) = 2\pi I_0[p N/(4 \pi)]$, where
$\rho_{\rm max}=N/(2\pi)$ is the radius of a perfect circle of
perimeter $N$.  (The value of $\rho_{\rm max}$ can also be obtained
from the condition that $\int |\chi(\rho)|^2 d\rho \geq 1$, \ie that
going from the center of mass outward, one must cross the ring at
least once.)

Within the zeroth-order approximation, and in the thermodynamic limit,
the partition function is given by
\begin{equation}
  \label{eqTMZ}
  Z^{(0)}(p,N) = [\mu_{\rm max}^{(0)}]^N = [2 \pi I_0 (p N/4\pi)]^N.
\end{equation}
This result has a straightforward interpretation. As shown in the
Appendix [\Eq{eqZf}], it is identical to the partition function of a
2D, open, freely jointed chain subject to a tensile force
$f=pN/(4\pi)$. This force is just the tension associated with a
Laplace pressure $p$ acting on a circle of radius $\rho_{\rm max}\sim
N$.  The mean area is obtained from \Eq{eqTMZ} as
\begin{equation}
  \label{eqZerothOrderMeanArea}
  \langle A (\hat p,N)\rangle = \D{\ln Z}{p} = 
  \frac{N^2}{4\pi}\frac{I_1(\hat{p})}{I_0(\hat{p})},
\end{equation}
which saturates, as expected, to $\Amax=N^2/(4\pi)$ as $p\rightarrow
\infty$. The approach to saturation is given by
\begin{equation}
  \langle A(\hat p\gg 1)\rangle/\Amax \simeq 1 - \frac{1}{2\hat p},
\end{equation}
which corrects the mean-field prediction, \Eq{eqMeanAreaAtPc}, by a
factor of $2$.

\section{Monte Carlo Simulations}\label{secMC}

Since our Flory argument and mean-field theory may fail near the
critical point, we conducted Monte Carlo simulations to obtain the
mean area $\A$, mean-square radius of gyration $\RgSq$ and mean-square
area fluctuation $\langle \Delta A^2 \rangle$ as a function of
pressure $p$ for different ring sizes $N$.
\subsection{Numerical Scheme}
\begin{figure}[tbh]
  \centering
  \vspace{1in}
  \includegraphics[width=2.5in]{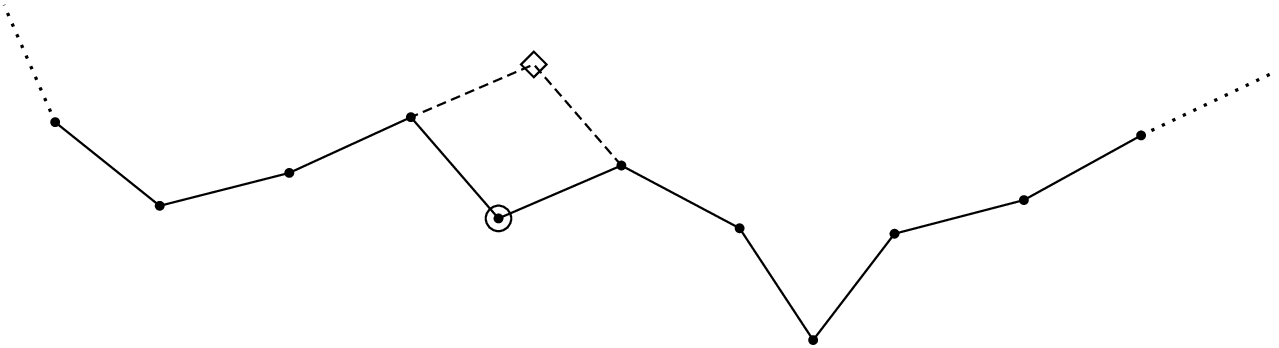}
  \vspace{0.2in}
  \caption[Schematic view of a section of the simulated polygon]{
    Schematic view of a section of the simulated polygon. A randomly
    chosen vertex (marked by a circle) can be moved to a single
    position only (marked by a diamond) so as to maintain the lengths
    of the two links attached to it constant. The area enclosed by the
    resulting rhomb is the difference in total polygon area for the
    given step.}
  \label{figPolygonSchema}
\end{figure}
We consider a polygon of $N$ equal edges, defined by the 2D
coordinates of its vertices. An off-lattice simulation is used, \ie
the positions of the vertices are defined in continuous space.  At
each step a random vertex is chosen and moved to the only other
position that satisfies the edge-length constraint
(\Fig{figPolygonSchema}). The difference in energy between the two
steps is proportional to the difference in total area, which in turn
is simply given by the area of the rhomb composed of the two edges
prior to and after the move (see \Fig{figPolygonSchema}). This way
each step takes only O(1) operations. The move is subsequently
accepted or rejected according to the Metropolis criterion.

The initial configuration is a stretched, regular polygon. This
initial condition and the dynamics defined above imply that the
polygon angles are restricted to change in discrete quanta of $\pm2\pi
/N$. Thus, although the algorithm is off-lattice, the simulated ring
is a discrete variant of a freely jointed chain which strictly
coincides with the freely jointed model only for $N\rightarrow\infty$.

The simulations were performed for $N$ between 50 and 3200, and for
$p$ between 0 and $4\Pc$. Away from the critical point, the number of
steps required for equilibration is $O(N^3)$, but near
$\Pc$ the simulation length must be extended due to critical slowing
down \cite{Binder1979}.  This limited our investigation of the
transition to $N \lesssim 3000$.

\subsection{Results}
\begin{figure}[tbh]
  \centering
  \vspace{0.4cm}
  \includegraphics[width=3.2in]{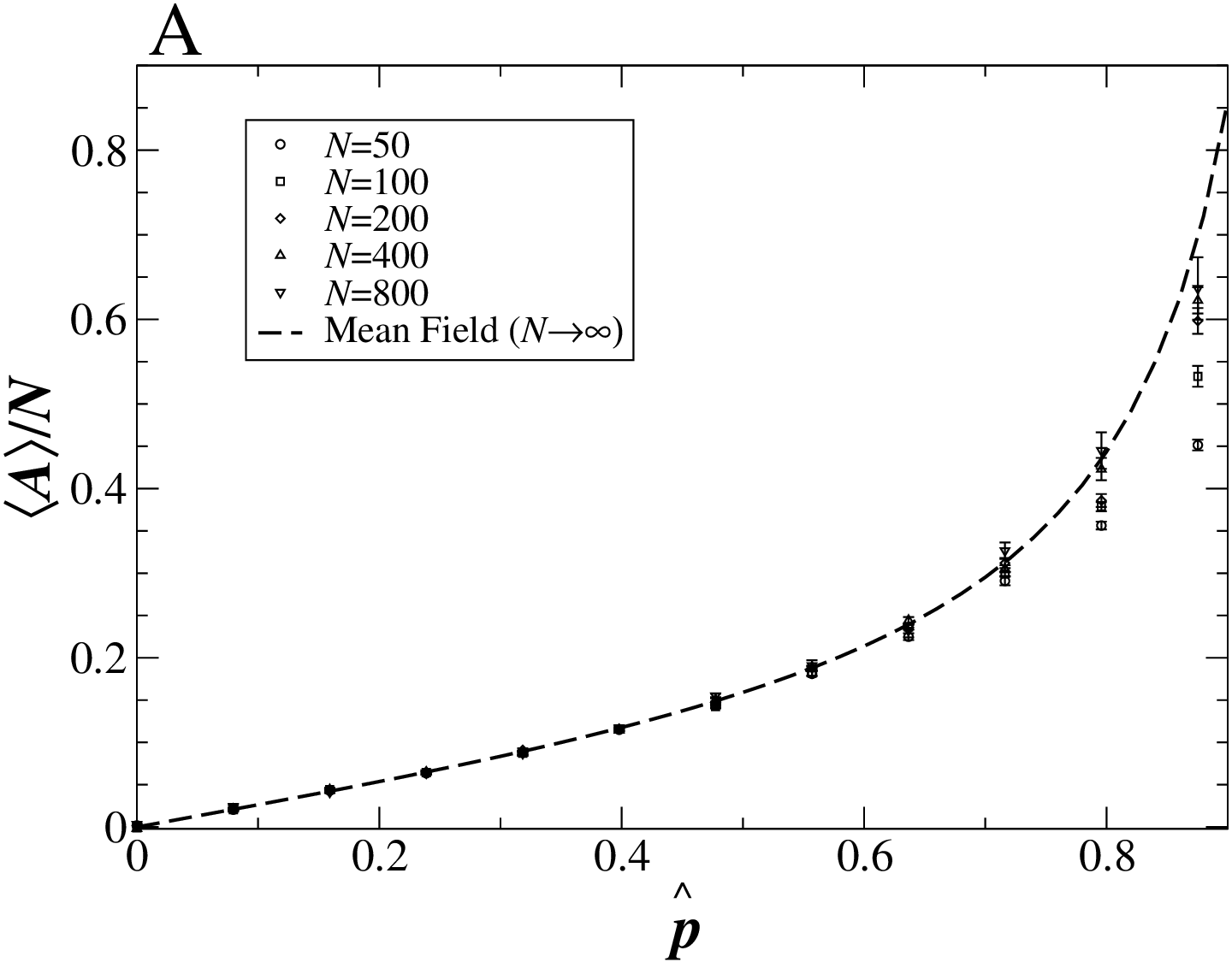}
  \vspace{0.6cm}
  \includegraphics[width=3.2in]{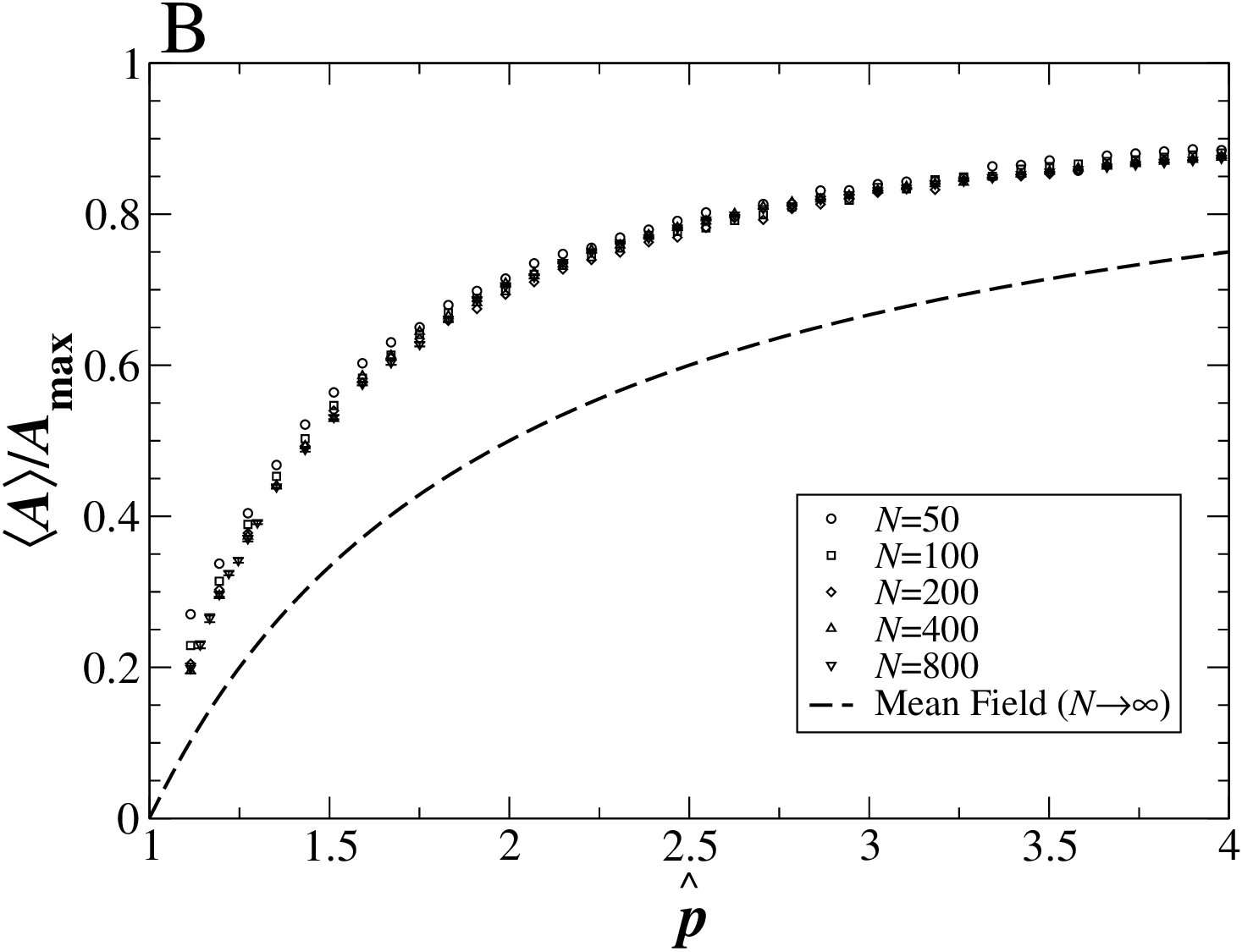}
  \caption{Mean area as a function of pressure below (A) and above (B)
    the critical point. The pressure is scaled by $\Pc=4\pi/N$, and
    the area by $N$ in (A) and by $A_{\rm max}=N^2/(4\pi)$ in
    (B). Symbols show the results of MC simulations for different
    values of $N$.  The dashed lines show the prediction of the
    mean-field theory in the limit $N\rightarrow\infty$.}
  \label{figCollapseOfData}
\end{figure}

Figure \ref{figCollapseOfData} shows simulation results for the mean
area as a function of pressure for different values of $N$. When $p$
is scaled by $\Pc\sim N^{-1}$ and $\A$ by $N$ (below $\Pc$) or by
$\Amax\sim N^2$ (above $\Pc$), the data below and above the transition
collapse onto two universal curves, thus confirming the predicted
scaling laws for the crumpled and smooth states (see Sections
\ref{secFlory} and \ref{secMF}). However, while the data well below
$\Pc$ coincide with the scaling function obtained from the mean-field
approximation, the data above the critical pressure collapse onto a
different curve.

\begin{figure}[tbh]
  \vspace{0.4cm}
  \centering
  \includegraphics[width=3.2in]{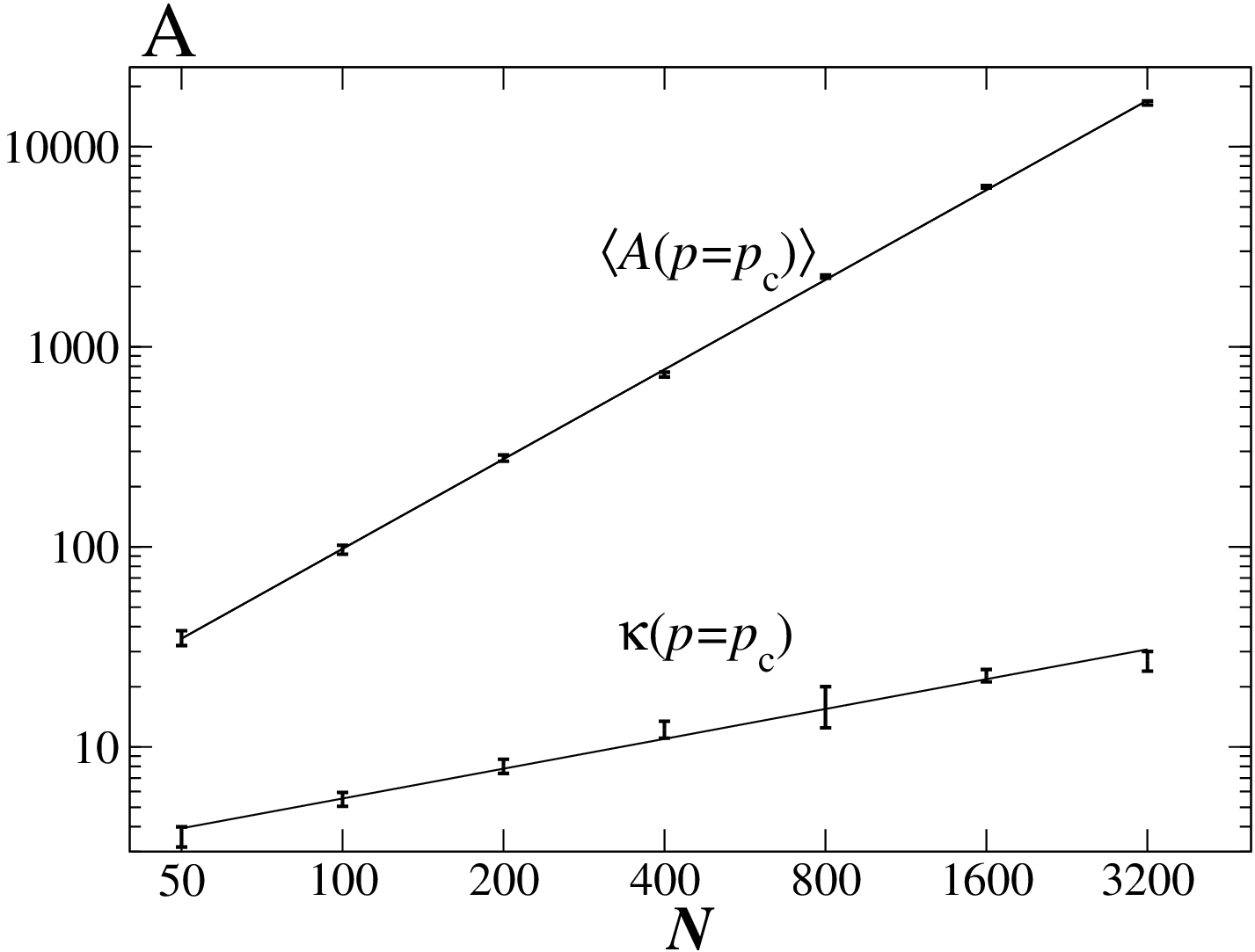}
  \vspace{0.6cm}
  \includegraphics[width=3.2in]{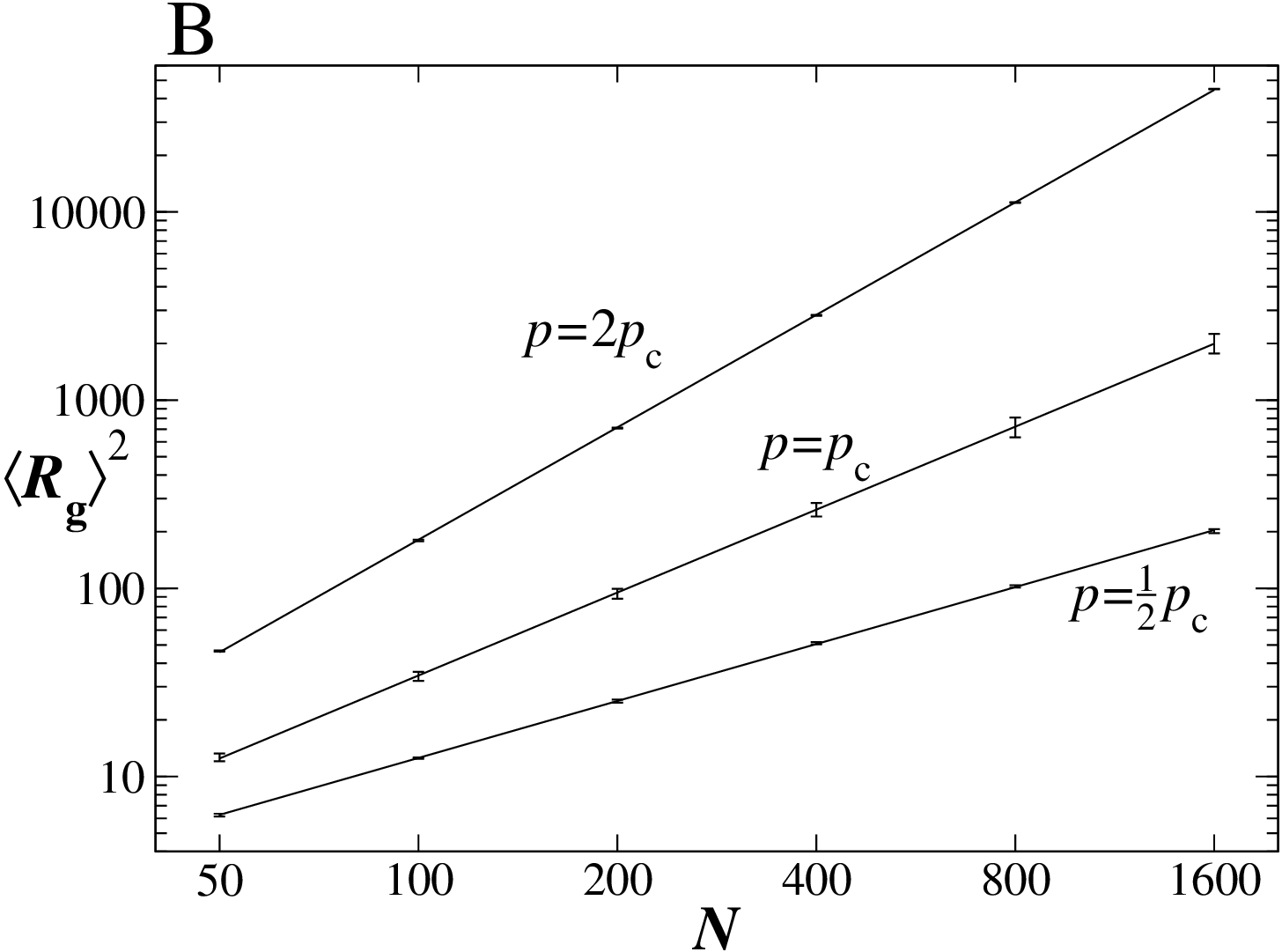}
  \vspace{0.2cm}
  \caption{ (A) Mean area and compressibility at $p=\Pc$ as a function
    of $N$ obtained by MC simulations. The fits, $\langle A(\Pc)\rangle
    = (0.102\pm 0.007)N^{1.49\pm 0.01}$ and $\kappa(\Pc)=(0.56\pm
    0.09)N^{0.495\pm 0.025}$, are given by the solid lines. (B)
    Mean-square radius of gyration as a function of $N$ at different
    values of $p=\Half\Pc$, $\Pc$ and $2\Pc$ with the best fits (solid
    lines) $\langle R_{\rm g}^2(\Half\Pc)\rangle=(0.12\pm0.01)
    N^{1.01\pm 0.02}$, $\langle R_{\rm g}^2(\Pc) \rangle= (0.043\pm
    0.025)N^{1.46\pm 0.13}$, and $\langle R_{\rm g}^2(2\Pc) \rangle=
    (0.019\pm0.001) N^{1.985\pm 0.011}$.  }
  \label{figCriticalExponents}
\end{figure}

The simulation results for the mean area and compressibility at
$p=\Pc$ as a function of $N$ are shown in
\Fig{figCriticalExponents}A. The reduced compressibility, defined in
\Eq{eqCompressibility}, was calculated from the measured mean-square
area fluctuation as $\kappa = (4\pi/N) \langle \Delta A^2
\rangle/\A$. We get
\begin{eqnarray}
  \label{eqMCAK}
  \langle A(p=\Pc)\rangle &= (0.102\pm 0.007)N^{1.49\pm 0.01}\\
  \kappa(p=\Pc) &= (0.56\pm 0.09)N^{0.495\pm 0.025}.
\end{eqnarray}
Hence, the mean-field exponents for the scaling with $N$,
\Eqs{eqMeanAreaAtPc}{eqMeanCompAtPc}, are confirmed. The predicted
prefactor of the compressibility is within the standard error of the
fit while that of the mean area differs by 4 standard errors.

We also measured from simulations the dependence of the mean-square
radius of gyration on $N$, as depicted in \Fig{figCriticalExponents}B.
As predicted, below the critical pressure we find $\RgSq\sim
N^{1.01\pm 0.02}$, and above it $\RgSq\sim N^{1.985\pm 0.011}$. At
$p=\Pc$ we get
\begin{equation}
  \label{eqMCRg}
  \langle R_{\rm g}^2(p=\Pc) \rangle= (0.043\pm 0.025)N^{1.46\pm 0.13}.
\end{equation}

\begin{figure*}[tbh]
  \centering
  \vspace{0.7cm}
  \mbox{\includegraphics[height=2.5in]{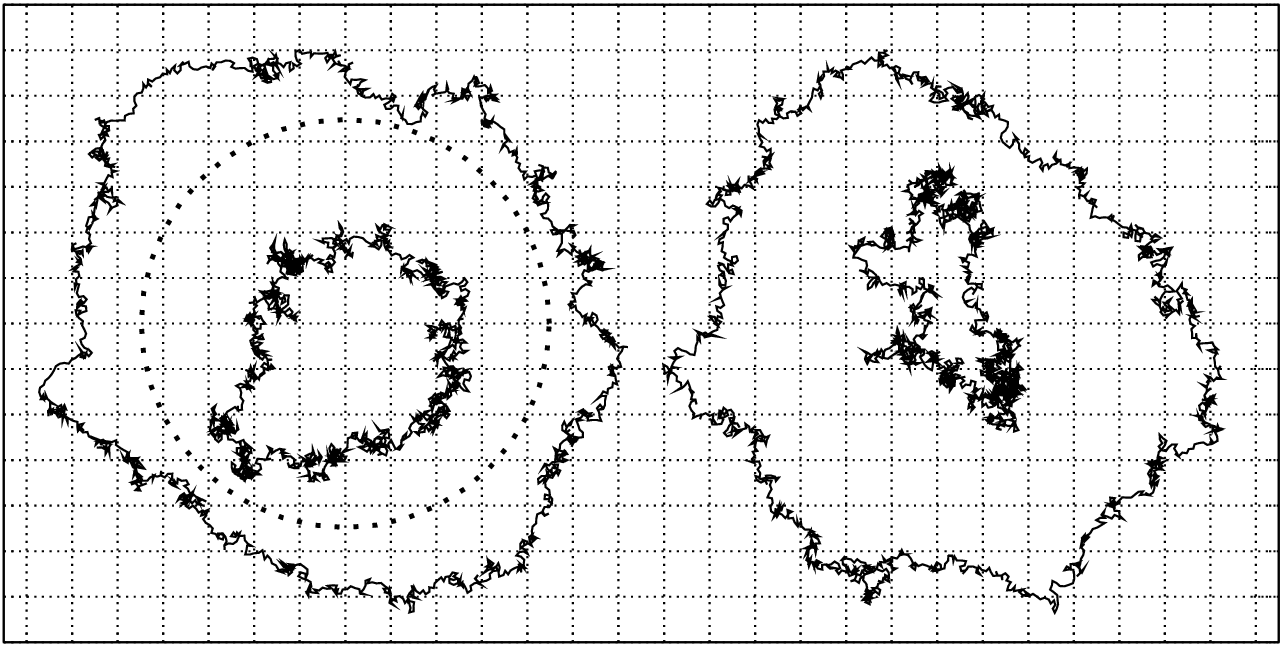}}
  \caption{ Four random conformations of a ring with $N=1600$ at the
    critical state as obtained by MC simulations, demonstrating the
    critical fluctuations. (The positions of the rings have no
    significance.)  The spacing between gridlines is ten times the
    link length. The dotted circle, shown for reference, has an area
    equal to the mean area of the ring at this state. }
  \label{figConformations}
\end{figure*}

For illustration we show in \Fig{figConformations} four randomly
chosen conformations of an 1600-segment ring at the critical pressure.
The shapes vary in size significantly due to critical fluctuations.

\begin{figure}[tbh]
  \centering
  \vspace{0.2in}
  \includegraphics[width=3.2in]{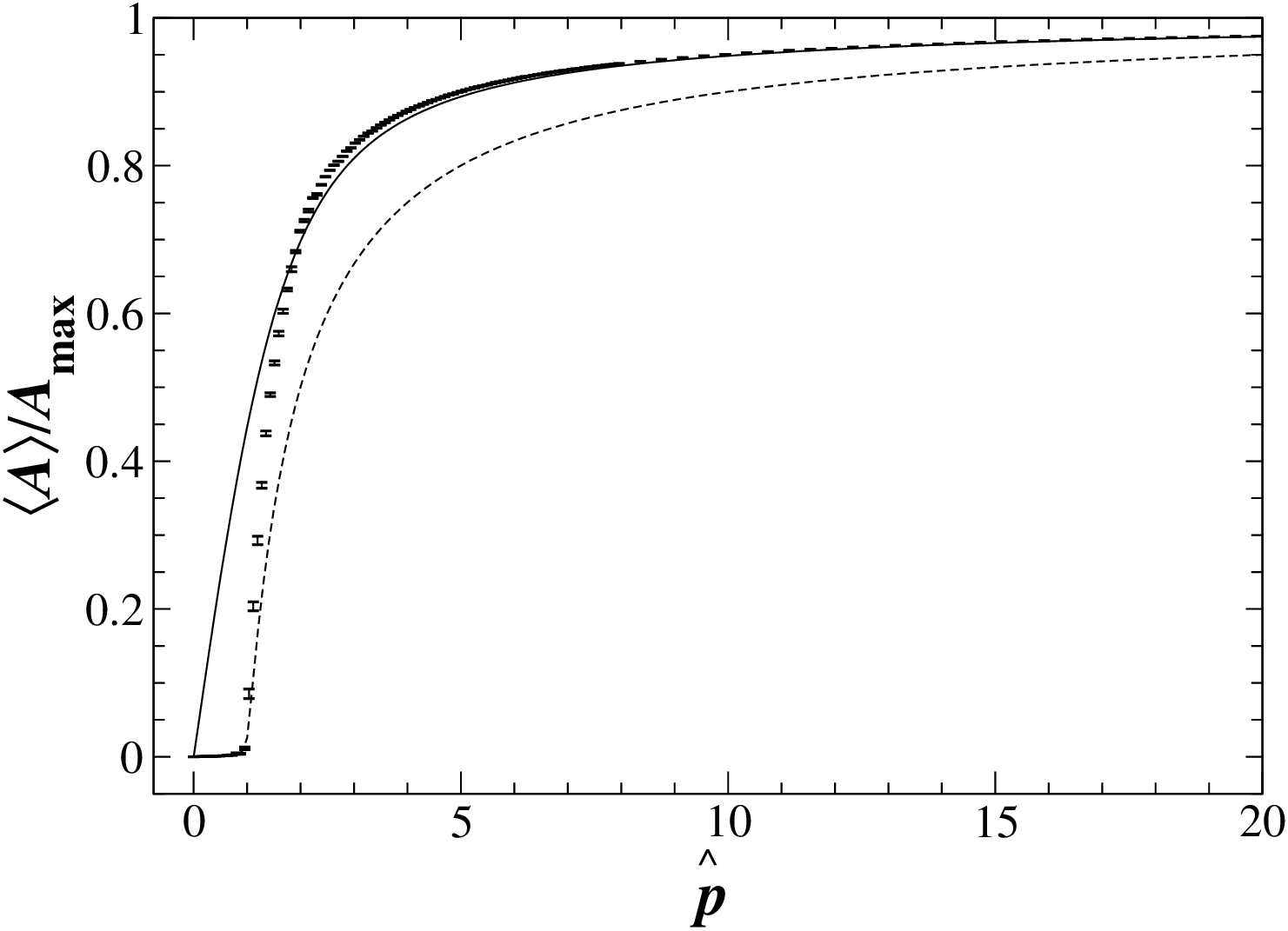}
  \caption{ Mean area (scaled by $A_{\rm max}$) as a function of
    pressure (scaled by $\Pc$), as obtained by the zeroth-order
    transfer-matrix calculation (solid curve), mean-field
    approximation (dashed curve), and MC simulations for $N=1600$
    (error bars).}
  \label{figAreaFromAll}
\end{figure}

The simulation results for the smooth state, $p>\Pc$, are shown in
\Fig{figAreaFromAll}, where they can be compared with the
transfer-matrix calculation, \Eq{eqZerothOrderMeanArea}, and the
mean-field result, \Eqs{eqMeanUSquare}{eqMeanArea2}. On the one hand,
there is good agreement with the transfer-matrix calculation for
$\hat{p}\gtrsim 5$, particularly compared to the mean-field result. On
the other hand, the mean-field theory succeeds in reproducing the
phase transition, whereas the zeroth-order transfer-matrix calculation
is invalid for these low pressures.

\subsection{Critical Exponents} \label{secCE}

We have confirmed the predicted exponents relating $\A$, $\RgSq$ and
$\kappa$ at $p=\Pc$ with the system size $N$. We now turn to the
exponents characterizing the divergence with $|\hat{p}-1|$.  As in any
continuous phase transition, the critical fluctuations make it hard to
accurately measure these critical exponents from simulations. Instead,
we choose the more reliable route of finite-size scaling to obtain the
relations between them.

Let us divide the ring at the critical state into Pincus correlation
blobs \cite{PolymerPhysics2003,Pincus1976} of $g$ monomers and
diameter $\xi$ each, such that within a blob the polymer behaves as an
unperturbed random chain, \ie $g\sim\xi^2$, whereas the chain of blobs
is stretched by the pressure. The perimeter of the ring is
$R\sim(N/g)\xi\sim N/\xi$.  We have already established that the
perimeter of the ring at $p=\Pc$ scales as $R\sim N^{3/4}$. (See
\Fig{figCriticalExponents}B.) Thus, for a finite-size system, we get a
correlation length which scales at $p=\Pc$ as
\begin{equation}
\label{eqXi}
  \xi(p=\Pc) \sim N^{1/4}.
\end{equation}
On the other hand, close to the critical point the correlation length
diverges as $\xi\sim |\hat{p}-1|^{-\nu}$, the compressibility as
$\kappa\sim |\hat{p}-1|^{-\gamma}$, and the order parameter increases
as $M\sim |\hat{p}-1|^{\beta}$. Using \Eq{eqXi} and the numerically
established results, $M=R/N\sim N^{-1/4}$ and $\kappa\sim N^{1/2}$, we
obtain the relations
\begin{equation}
  \beta=\nu,\ \ \ \gamma=2\nu,
\end{equation}
which hold for the mean-field universality class. (Note that the
exponents $\alpha$ and $\delta$ are irrelevant for this system, since
both the ordering field and temperature are incorporated in the single
parameter $p$.)

\section{Discussion}\label{secSummary}

We have demonstrated that the swelling of a 2D freely jointed ring due
to a pressure differential exhibits a second-order smoothening
transition of the mean-field universality class.  Below the critical
pressure the ring behaves as a random walk, with both the mean area
$\A$ and mean-square radius of gyration $\RgSq$ proportional to $N$.
In this crumpled state the mean area obeys the scaling
$\A=Nf_A^<(p/\Pc)$. Mean-field theory accurately captures the scaling
law as well as the scaling function $f_A^<$.  (See
\Fig{figCollapseOfData}A.) This lies in the fact that, for an
unstretched chain, the Gaussian-spring description in the mean-field
calculation and the actual freely jointed model coincide as
$N\rightarrow\infty$.

Above the critical pressure $\A$ and $\RgSq$ are proportional to
$N^2$.  For this smooth state we have found a new scaling behavior,
$\A=N^2f_A^>(p/\Pc)$ (\Fig{figCollapseOfData}B). The mean-field theory
correctly gives the scaling law, yet fails to predict the correct
scaling function $f_A^>$.  This is because in a stretched state the
entropy of a chain of Gaussian, variable springs is much larger than
that of a freely jointed chain of rigid links. For $p\gg\Pc$ we have
calculated $f_A^>$ exactly [\Eq{eqZerothOrderMeanArea} and
\Fig{figAreaFromAll}].

At the transition point itself a new state of polymer chains has been
discovered, with $\A$ and $\RgSq$ proportional to $N^{2\nu_{\rm c}},
\nu_{\rm c}=3/4$. The swelling exponent $\nu_{\rm c}$ turns out to be
identical to that of a 2D self-avoiding walk, although the physical
origins of the two exponents are unrelated; the freely jointed ring in
this intermediate state between crumpled and smooth behaviors contains
numerous intersections, as is clearly seen in \Fig{figConformations}.

Our discussion has been restricted to freely jointed,
self-intersecting chains. We have commented in Section \ref{secFlory}
that the addition of self-avoidance to the Flory free energy removes
the second-order transition. Thus, the current work suggests a
different perspective for the previously studied swelling behavior of
2D self-avoiding rings \cite{Fisher1987,Fisher1990}. The
large-inflation scaling regime thoroughly discussed in those works can
be viewed as a broadening of a phase transition present in the
non-self-avoiding model, the self-avoidance thus acting as a relevant
parameter for the critical point discussed in the current work.

One of the strongest assumptions underlying our analysis, as well as
those of Refs.\ \cite{Rudnick1991,Rudnick1993,Levinson1992}, is the
replacement of the actual geometrical area of the ring with its
algebraic area. An important issue is how this assumption affects our
results concerning the transition. It is clear that negative
contributions to the algebraic area are significant in the crumpled,
random-walk state and insignificant in the smooth state. The key
question, therefore, is whether they are statistically significant at
the transition. Returning to \Fig{figConformations} and the blob
analysis presented in Section \ref{secCE}, we infer that the negative
area contributions lie within the correlation blobs. We have
shown that each blob contains $g\sim \xi^2\sim N^{1/2}$ monomers. The
deviation of the geometrical area of a blob from its algebraic area is
$\langle a^2\rangle^{1/2}\sim \xi^2\sim N^{1/2}$. (Recall that the
blob contains an unperturbed chain with zero mean algebraic area.)
The ring contains $n=N/g\sim N^{1/2}$ such blobs. Hence, the total
deviation of the geometrical area from the algebraic one is $(n\langle
a^2\rangle)^{1/2}\sim N^{3/4}$.  In the limit $N\rightarrow\infty$
this is negligible compared to the mean area of the ring at the
critical state, $\A\sim N^{3/2}$. Thus, we conjecture that the same
smoothening transition as the one reported here will be found also in
a model which considers the geometrical area rather than the algebraic
one.

\footnotesize{We thank M.\ Kroyter, H.\ Orland, P.\ Pincus and T.\
  Witten for helpful discussions.  This work was supported by the
  US--Israel Binational Science Foundation (Grant no.\ 2002-271).}

\section*{Appendix: 2D Freely Jointed Chain under Tension}

\setcounter{equation}{0}
\renewcommand{\theequation}{A.\arabic{equation}}

In this Appendix we recall the results for the partition function of
a 2D, {\it open}, freely jointed chain subject to a tensile force. These
results are used in Sections \ref{secFlory} and \ref{secTM}.

Consider a 2D freely jointed chain composed of $N$ links of length
$l\equiv 1$.  A chain configuration is defined by $N$ 2D unit vectors,
$\{{\bf u}_j\}_{j=1,\ldots,N}$, specifying the orientations of the
links.  One end of the chain is held fixed while the other is pulled
by a force ${\bf f}$ (measured in units of $\KT/l$). The partition
function of the chain is
\begin{equation}
\label{eqZf}
\begin{split}
  Z({\bf f},N) &= \int \prod_{j=1}^N d{\bf u}_j e^{{\bf f}\cdot {\bf u}_j} \\
  &= \left( \int_0^{2\pi} d\theta e^{f\cos\theta} \right)^N = [2\pi I_0(f)]^N.
\end{split}
\end{equation}
  
The mean end-to-end vector is obtained from \Eq{eqZf} as
\begin{equation}
{\bf R} = \nabla_{\bf f}\ln Z = N\frac{I_1(f)}{I_0(f)}{\bf\hat f}.
\end{equation}
The mean end-to-end distance in the limit of weak force, to two
leading orders, is
\begin{equation}
  R/N \simeq f/2 - f^3/16,
\end{equation}
leading, upon inversion, to
\begin{equation}
  f \simeq 2R/N + (R/N)^3.
\end{equation}
Finally, the free energy for fixed end-to-end distance (to two leading orders
in small $R$) is
\begin{equation}
  \label{eqApF}
  F(R)=-\ln Z[{\bf f}({\bf R})]+{\bf f}\cdot{\bf R}\simeq\frac{R^2}{N}+
  \frac{R^4}{4N^3}.
\end{equation}
This yields the usual Gaussian term, $F_{\rm el}$, and the first
correction due to inextensibility, $F_{\rm inext}$, used in
\Eq{eqFloryEnergy}.

\bibliographystyle{unsrt}

\end{document}